\documentclass[runningheads]{llncs}
\usepackage{graphicx}
\usepackage{xcolor}
\usepackage{caption}
\usepackage{multirow}
\usepackage{subcaption}
\usepackage{mathtools}
\usepackage{amsmath,amssymb,amsfonts}

\usepackage{amsmath,amssymb}
\usepackage{lipsum}
\usepackage{algorithm}
\usepackage[noend]{algpseudocode}
\begin{document}
\title{Consensus-based Participatory Budgeting for Legitimacy: Decision Support via Multi-agent Reinforcement Learning}
\titlerunning{Consensus-based Participatory Budgeting for Legitimacy}
%
\author{Srijoni Majumdar\inst{}\and
Evangelos Pournaras\inst{} }
%
%
\institute{School of Computing, University of Leeds, Leeds, UK}
%
\maketitle              
\begin{abstract}

 The legitimacy of bottom-up democratic processes for the distribution of public funds by policy-makers is challenging and complex. Participatory budgeting is such a process, where voting outcomes may not always be fair or inclusive. Deliberation for which project ideas to put for voting and choose for implementation lack systematization and do not scale. This paper addresses these grand challenges by introducing a novel and legitimate iterative consensus-based participatory budgeting process. Consensus is designed to be a result of decision support via an innovative multi-agent reinforcement learning approach. Voters are assisted to interact with each other to make viable compromises. Extensive experimental evaluation with real-world participatory budgeting data from Poland reveal striking findings: Consensus is reachable, efficient and robust. Compromise is required, which is though comparable to the one of existing voting aggregation methods that promote fairness and inclusion without though attaining consensus.

\end{abstract}

\section{Introduction}

Participatory budgeting (PB) is a bottom-up collective decision-making process with which citizens decide how to spend a budget of the local municipality~\cite{wellings2023improving,faliszewski2023participatory}. Citizens initially submit proposals for implementation of various project ideas, i.e. public welfare amenities. These are evaluated by the city officials and finally, a subset is put for voting. Citizens then express their preferences using different input voting methods such as approval or score voting~\cite{kilgour2010approval}. Finally, voting aggregation methods are applied to select the winner projects~\cite{faliszewski2023participatory}. 

The selection of the winner projects depends on both input and aggregation methods~\cite{aziz2021participatory}. As preferences via approvals or scores are based on self-interest, voting outcomes may yield different satisfaction levels, under-representation, and poor legitimacy. For a more stable, conclusive, shared and legitimate voting outcome, a form of systematic and scalable deliberation is missing among citizens so that individual preferences are exchanged, debated and compromised in a viable way to reach consensus~\cite{ganuza2012deliberative}. This challenge is addressed in this paper. 

A new multi-agent reinforcement learning approach (MARL-PB) is introduced to model a novel iterative consensus-based PB process. In the proposed approach, consensus emerges as a result of (i) reward-based learning based on project ideas proposed and selected in the past and (ii) decentralized voter communication that supports information exchange and deliberation. 

\begin{figure}[!htb]
    \centering
    \includegraphics[scale = 0.17]{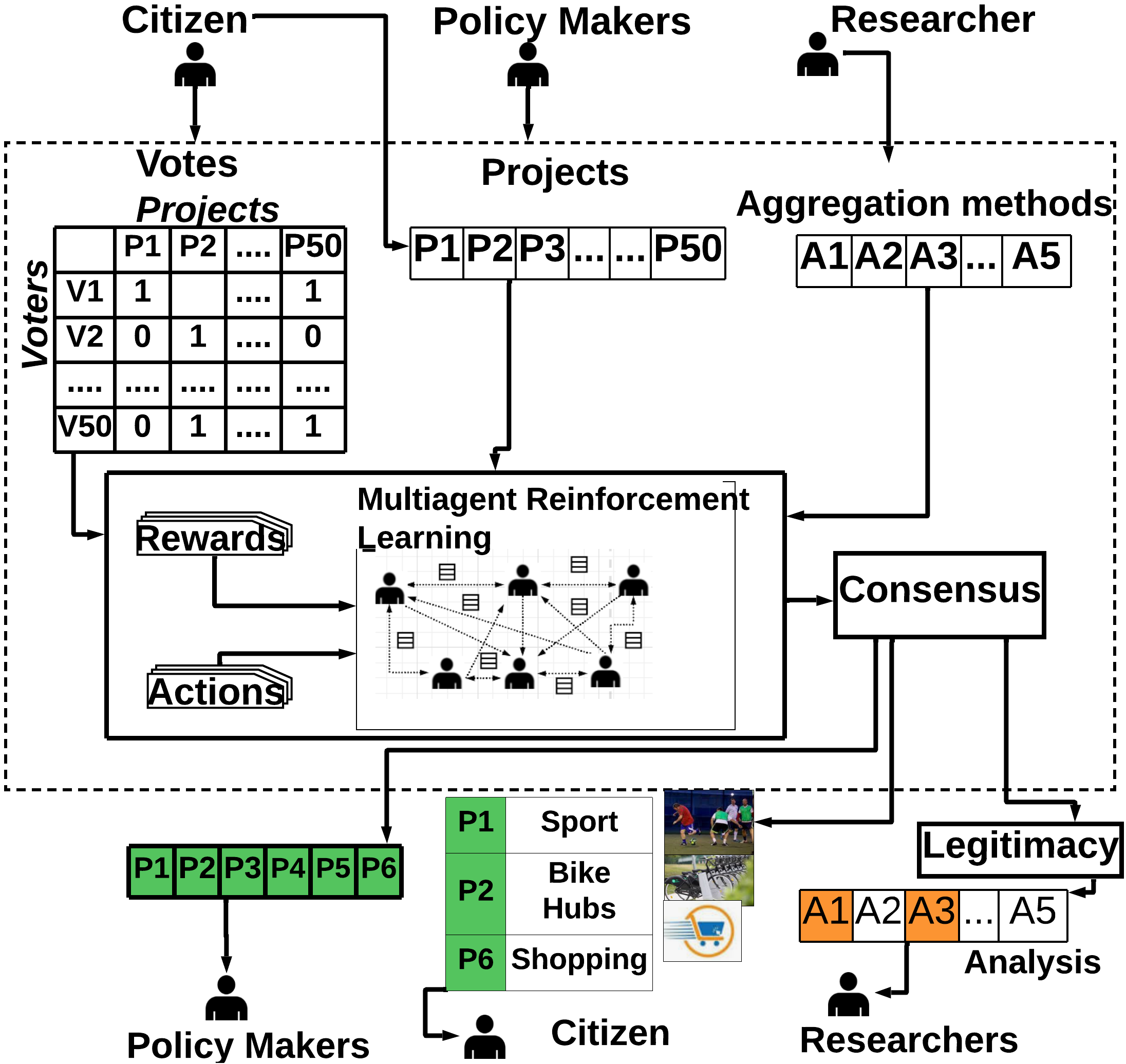}
    \caption{Consensus-based participatory budgeting using a multi-agent reinforcement learning (MARL-PB). A decision-support framework is designed for three different use cases and beneficiaries: citizens, policy-makers and researchers.}
    \label{fig:archi}
\end{figure}

MARL-PB is implemented as a decision-support system that finds applicability in three use cases by three beneficiaries as shown in Figure~\ref{fig:archi}: (i) \emph{Citizens}: digital assistance to communicate, deliberate and reach a common ground for which projects to implement. This is expected to increase the participation, satisfaction and legitimacy in participatory budgeting. (ii) \emph{Policy-makers}: digital assistance to filter out projects during the project ideation phase with the aim to put for voting a reasonable and legitimate number of projects that results in informed and expressive choices during voting without informational overload. (iii) \emph{Researcher}: digital assistance for the assessment of fair and inclusive voting aggregation methods (e.g. equal shares, Phragmen) via comparisons with a fine-grained consensus-based model such as the one of MARL-PB.  

MARL-PB is extensively assessed using state-of-the-art real-world participatory budgeting datasets~\cite{stolicki2020pabulib} from Poland. The following three research questions are addressed: (i) \emph{How effective multi-agent reinforcement learning is to assist voters reach consensus in participatory budgeting?} (ii) \emph{What level of flexibility is required by voters to compromise and reach consensus in participatory budgeting?} (iii) \emph{How efficient and robust a consensus-based participatory budgeting is by using multi-agent reinforcement learning?}. The quality of consensus, its efficiency and robustness are studied, along with how they are influenced by factors such as the following: (i) number of possible consensus bundles, (ii) in-degree of the communication network, (iii) number of voters, (iv) districts, (v) voting aggregation methods and (vi) project attributes.

The contributions of this paper are summarized as follows: (i) The multi-agent reinforcement learning approach of MARL-PB to model and implement an iterative consensus-based PB process. (ii) A decision-support framework based on MARL-PB to digitally assist three use cases by three beneficiaries. (iii) An extension of the reward-based learning strategy with a gossip-based agents communication protocol for decentralized information exchange and consensus building. (iv) A compilation of metrics that characterize and assess the legitimacy of the consensus-based PB process. (v) Practical and revealing insights about the nature of the achieved consensus: requires compromises comparable to the ones of the voting aggregation methods that promote fairness and inclusion. (vi) An open-source software artifact of MARL-PB for reproducibility and encouraging further research in this niche research area {\tiny \footnote{ https://github.com/DISC-Systems-Lab/MARL-PB (last accessed: July 2023).}}

This paper is outlined as follows: Section~\ref{sec:related-work} reviews related work. Section~\ref{sec:Methodology} introduces the consensus-based approach. Section~\ref{sec:Results} illustrates the empirical results and findings. Section~\ref{sec:Conclusion} concludes this paper and outlines future work.

\section{Related Literature Review}\label{sec:related-work}

This section provides an overview of related literature, with a focus on iterative reward-based learning for collective decision-making processes.

Social dilemma games such as Prisoners Dilemma has been studied in the context of reward-based learning agents~\cite{macy2002learning} with two agents and discrete rewards, i.e. punishment (0) or no punishment (1), to explore the compromises that two agents make to reach consensus. Using multiple agents, this experiment provides insights on how learning can stabilize using limited voters and deterministic rewards. This provides a relevant direction for dealing with voting for social choice and collective preferences. 

Airiau et al.~\cite{airiau2017learning} model an iterative single-winner voting process in a reinforcement learning setup to analyze the learning capabilities of voting agents to obtain more legitimate collective decisions.  The proposed framework provides a new variant of iterative voting that allows agents to change their choices at the same time if they wish.  The rank of the winner at every stage in the preference order of voters is used as a reward for the agents to learn and re-select. The proposed work by Liekah et al.~\cite{liekah2019multiagent} additionally calculates the average satisfaction among voters in every iteration based on the winner and individual preferences. 

\begin{table}[!htb] 
\caption{Comparison of this work with earlier multi-agent reinforcement learning approaches for collective decision making.}\label{tab:literature} 
\centering 
\resizebox{\columnwidth}{!}
{ \scriptsize \begin{tabular}{lllll}
\hline
Aspect & {Macy et al.~\cite{macy2002learning}} &{Airiau et al.~\cite{airiau2017learning}} & {Liekah et al.~\cite{liekah2019multiagent}} & {Proposed Approach (MARL-PB)} \\  \hline
{\bf Outcome} & Single Winner & Single Winner & Single Winner & Multiple Winners\\ 
{\bf Rewards} &  Deterministic& Stochastic$*$ & Stochastic$*$ & Deterministic (project attributes) \\
 &    4 discrete values & & &  Stochastic (from communication)\\
{\bf Execution} & Centralized & Centralized & Centralized & Shared aggregate rewards, decentralized \\
{\bf Action Space} & 4  & 5  & 5 & till 100 \\  \hline
\multicolumn{5}{l}{$*$Rank of  winner in the preference order of the voter (within an iteration).}\\
\end{tabular}} \vspace{-0.2cm}\end{table}

Prediction of complete PB ballots using machine learning classification is recently studied as a way to decrease information overload of voters using partial ballots~\cite{Gilla2023}. This approach could complement MARL-PB to speed up the consensus process. 

Existing approaches (see Table~\ref{tab:literature}) do not incorporate inter-agent communication for large-scale information exchange in multi-winner voting systems. The rewards are fixed in centralized settings and do not model the preferences of the voters. Moreover, the feasibility of reaching a consensus via communication with other voters has not been studied. This is relevant to the problem of scaling up and automating deliberation in collective decision-making to reach more legitimate voting outcomes. These are some of the gaps addressed in this paper. 

\section{Consensus-based Iterative Participatory Budgeting}\label{sec:Methodology}

In this section, an iterative participatory budgeting process is introduced modeled by a multi-agent reinforcement learning approach. The voters (agents) maximize their self-interest but also compromise to reach a consensus in a multi-agent system, where the choices of others are initially only partially known. 

\subsection{Multi-armed Bandit Formulation}

\begin{figure}[!htb]
    \centering
    \includegraphics[scale = 0.36]{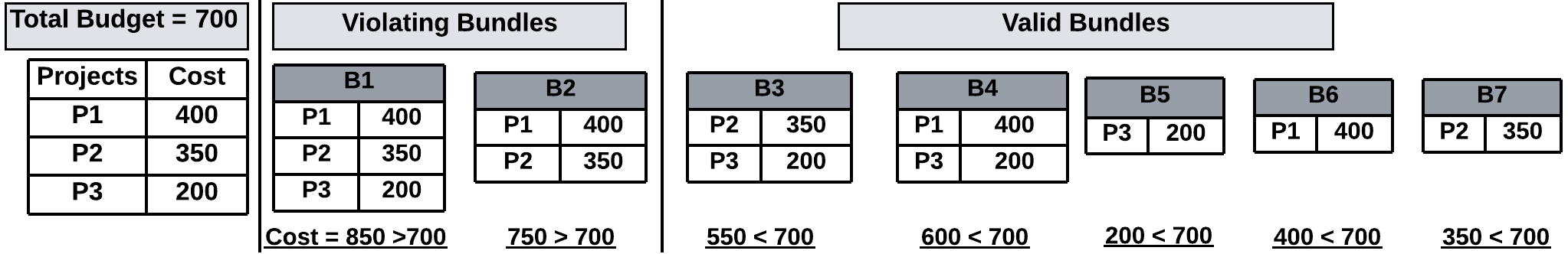}
    \caption{\textbf{Calculation of bundles}: They represent all possible combinations from the listed projects where the total cost of all the projects in the bundle is within the budget. In the example, three projects with their corresponding costs are listed for voting in a participatory budgeting process. The total budget is 700.  Five out of the seven possible bundles are valid and satisfy the budget constraint.}
    \label{fig:bundle}
\end{figure}

In a participatory budgeting process, voters collectively choose multiple projects subject to a constraint that the total cost of the projects is within the total budget. To incorporate this knapsack constraint, a combinatorial model~\cite{aziz2021participatory} is designed to formulate bundles from the available list of projects.  So for three projects, there are seven possible bundles, out of which a subset fulfills the budget constraints. These are referred to as {\em valid knapsack bundles} and they constitute the possible actions in a multi-arm bandit formulation (see Figure~\ref{fig:bundle}).

\vspace{1mm}

The bundles encode all possible multi-winner preferences the voters can collectively have. Learning valid knapsack bundles instead of individual project selections prevents early terminations by budget violations~\cite{badanidiyuru2018bandits}.

The iterative version of the PB process introduces a partial voters' communications at every iteration to exchange preferences with the aim they converge (compromise) to the same bundle (same preferences). This process models a large-scale automated deliberation process. The selection of an action by a voter depends only on the rewards associated with the bundles, hence, the problem is modeled as a multi-armed bandit reinforcement learning approach~\cite{slivkins2019introduction}.

The multi-arm bandits are defined in the form of a tuple $< A;R >$, where

\noindent $A$ represents actions that are the possible bundles and $R_A = P[r|A]$ is the probability distribution of rewards over the bundles (actions). 

{\em Actions:} For a participatory budgeting process with a set of projects and a total budget, the actions comprise of the {\em valid knapsack bundles} formed from the projects and their associated costs.

{\em Rewards}: The preference modelling in the form of rewards plays an important role in reaching consensus in a large action space with multiple agents. The aggregate preferences of the voters over past and the current year are encoded for a region in the form of rewards that signify how the needs for public amenities evolve over the years and thus can help to predict the collective preference for the current participatory budgeting process. These are modeled and calculated as deterministic rewards for each bundle. 

To reach a consensus, voters explore the action space of each other via information exchange. This exchange models a large-scale and automated deliberation process, which voters use to learn, compromise and adjust their choices.

\noindent{\em Deterministic Rewards}: A project is  related to a type of public welfare amenities\footnote{It is assumed that preferences for such projects persist over the passage of time, in contrast to infrastructure projects that once they are implemented, they may not be preferred anymore.} such as urban greenery, sports, culture, education, environmental protection etc., or a population group that benefits such as elderly, families with children, etc. The preference of citizens are mostly associated with these attributes and can be used to estimate collective preferences for the population of a region.

The number of occurrence of such project attributes, which are put for voting and selected in the past years of a region is used as reward utilities:

$$R^{a}\ =\ \Sigma_{y \in Y}  (\mathbb{C}(a) + \mathcal{C}(a)),$$

\noindent where $a$ is a specific project attribute, $\mathbb{C}$ and $\mathcal{C}$ signify the normalized total count of occurrence of the project attribute across listed and selected projects respectively over $Y$ years of participatory budgeting processes in a region.

The reward for a project is determined as follows:

$$R_{p} = \sigma( \Sigma_{i=1}^{\mathcal{A}} (R^{a}_i)) + \tanh(\frac{c_p}{\mathcal{B}}), $$

\noindent where $\mathcal{A}$ is the total number of attributes associated with a project, $c_p$ is the individual project cost and $\mathcal{B}$ is the total budget of the PB process. 
The rewards for a bundle is the sum of the rewards of each of its projects. 

\noindent{\em Rewards from inter-agent communication:} At every iteration, we update a dynamic random bidirectional graph using a decentralized process such as the gossip-based peer sampling~\cite{jelasity2007gossip} for peer-to-peer communication. At each iteration, the connected agents send the bundle that has received the highest rewards (other randomized schemes are supported by the code), together with the reward itself. As the neighbors are randomly decided, the accumulated rewards from information exchange are stochastic. The stateless variant of the Q learning approach is augmented to incorporate rewards obtained from information exchange:

$$   Q(b_t) \leftarrow Q(b_t) + \alpha (r + \delta (\underset{\ \ \ \ b^c_t}{max} (Q (b_t^c) - Q(b_t))),$$ 

\noindent where $Q(b_t^c)$ is the rewards obtained via agent communication for a bundle b at time t. The introduced learning rate for rewards from information exchange -$\delta$ is set empirically to 0.1. The discount factor $\gamma$ in the Q learning is set to zero as future rewards are not considered. Algorithm~1 outlines the learning process. 

\begin{algorithm}[!htb]
\caption{Augmented Q-Learning for consensus in participatory budgeting.}
\label{CHalgorithm}
\begin{algorithmic}[1]
\State Populate project list and cost for the current participatory budgeting process
\State Initialize the fixed rewards for projects
\State Calculate rewards for all valid bundles
\For{each iteration $i$ $\geq$ $1$ }
\For{each voter $v$ $\in$ V}
\If{$i$ == 1} 
\State Assign the bundle with highest overlap to original individual preferences
\EndIf
\State Update random graph via the peer sampling service
\State Aggregate rewards of bundles from neighbors
\State Update total rewards for a bundle in the Q table
\State Select action (bundle) according to  $\epsilon$-greedy policy, $\epsilon$$\in$[0,1]
\EndFor
\EndFor
\end{algorithmic}
\end{algorithm}

Initially, the agents select a bundle according to their preference (first iteration) and then they start communication with other agents during which they adjust their preferred bundle. The selection of the bundle at each iteration is based on the cumulative sum of both rewards, which the agents maximize using an $\epsilon$ greedy exploration strategy. For a low number of projects and voters size, the initial preferences from the multi-winner approvals of the voters may result in a reduced action space for exploration.

\subsection{Assessment Model for Consensus}\label{sec:metrics}

The following metrics are designed to assess the quality of the consensus (legitimacy) based on popularity, representation and budget utilization that can increase the satisfaction of the citizens, increase participation  and improve the quality of the overall PB process~\cite{aziz2021participatory,miller2019modes}. The level of compromise made by voters is assessed. These metrics also characterize how difficult it is to reach a consensus in an iterative voting process for participatory budgeting. The metrics that model the legitimacy are outlined as follows: 

\begin{itemize}

\item {\em Compromise Cost}: The mean non-overlap (1 - mean overlap) of projects between the preferred bundle of the voters and the consensus bundle, calculated using the Jaccard Index~\cite{fletcher2018comparing}.

\item {\em Unfairness}:  The coefficient of variation of the {\em compromise cost} over all agents. 

 \item {\em Popularity} (fitness of consensus):  The normalized ranking score of the projects in the consensus bundle, calculated using the number of votes of each project from the original voters' preferences.

 \item {\em Budget Utilization}: The cost of the projects in the consensus bundle divided by the total available budget in the participatory budgeting process. 

\end{itemize}

\section{Experimental Evaluation}\label{sec:Results}

This section illustrates the results obtained from the evaluation of the consensus-based participatory budgeting process, using real-world data. These results shed light on the the efficiency of reward models, the communication protocol and the exploration strategy to reach consensus.

{\em Dataset:} 
The {\em pabulib} PB dataset 
(\url{http://pabulib.org/}) is used for the evaluation. It contains the metadata related to projects and voters along with the voting records for multiple participatory budgeting instances, for various districts and cities of Poland. Each project is associated with multiple attributes such as urban greenery, education, relavance to children etc., along with information about the project costs. There are multiple participatory budgeting instances for every district or city for multiple years and different ballot designs such as k-approval, cumulative and score voting. Furthermore, the winners are calculated using various aggregation methods such as the method equal shares, phragmen, and utilitarian greedy~\cite{faliszewski2023participatory} to assess the quality of the consensus bundles compared to the ones calculated by methods that promote fairness and inclusion. Three districts are selected - Ruda, Ursynow and Rembertow, whose valid bundles vary from a smaller set (12 for Ruda) to a larger one (90 for Rembertow).

\begin{table}[!htb]
 \caption{{\bf Parameters for experimentation with each dataset}. The ranges signify that experiments are performed incrementally, for instance, 5, 6, 7,....90 for the \# of bundles selected randomly in {\em Rembertow}. The maximum number of valid bundles extracted from 20 projects is 90. The available data is for 4 years for each district. The experiments are performed for the latest year and the aggregate preference (rewards) are calculated using all years. The decay rate and learning rates are set to 0.1 after empirical investigation.}
 \centering
\begin{tabular}{lllll}
\hline
\multicolumn{1}{c}{\bf Dataset} & \multicolumn{1}{c}{\bf \# of Projects} 
 & \multicolumn{1}{c}{\bf \# of Bundles} &  \multicolumn{1}{c}{\bf In-degree }  & \multicolumn{1}{c}{\bf \# of Agents}  \\ 
  \\ \hline
Rembertow & 20 & 5 to 90  & 2 to 26 &  50 to 100\\ 
Ursynow  & 18 & 5 to 75  & 2 to 26 &  50 to 100\\ 
Ruda  & 10 & 3 to 12  & 2 to 10 &  50 to 100\\  \hline
\end{tabular}
\label{tab:set}
\end{table}

 {\em Design}: 
The framework is tested using various settings such as the numbers of combinations (bundles), number of agents, learning rate, decay rate, and the in-degree of the random graph updated at every iteration (see Table~\ref{tab:set}). For each of these settings,  the projects selected in the consensus are analyzed and compared with winners selected using other aggregation methods such as the method of equal shares and greedy~\cite{faliszewski2023participatory}.

\subsection{RQ1: Effectiveness to Reach Consensus}

\begin{figure}[!htb]
    \centering
    \includegraphics[scale = 0.49]{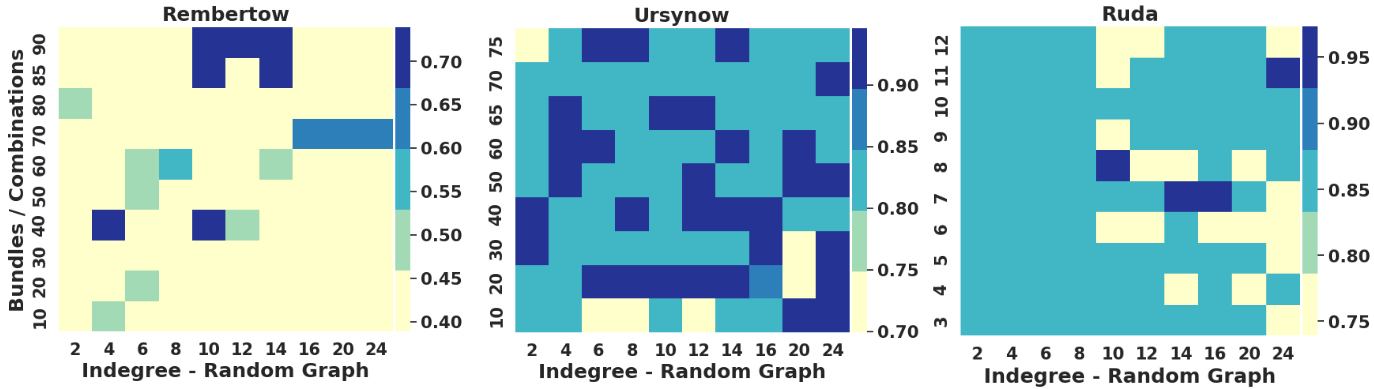}
    \caption{Total budget utilization of the consensus bundles for different number of valid bundle combinations and in-degree. The budget utilization for the consensus bundles for Ursynow is the highest.}
    \label{fig:budget}
\end{figure}

{\em Quality of Consensus}: The convergence to the consensus bundle depends on the in-degree of the random graph and the number of bundles (action space). Figure~\ref{fig:budget} shows the budget utilization as a function of the number of bundles and in-degree. Budget utilization of the consensus bundles (Figure~\ref{fig:budget}) for Rembertow is low (0.40 to 0.45) in most cases, which could be attributed to considerable high costs for popular projects and fewer projects in the consensus bundle. 

\begin{figure}[!htb]
    \centering
    \includegraphics[scale = 0.49]{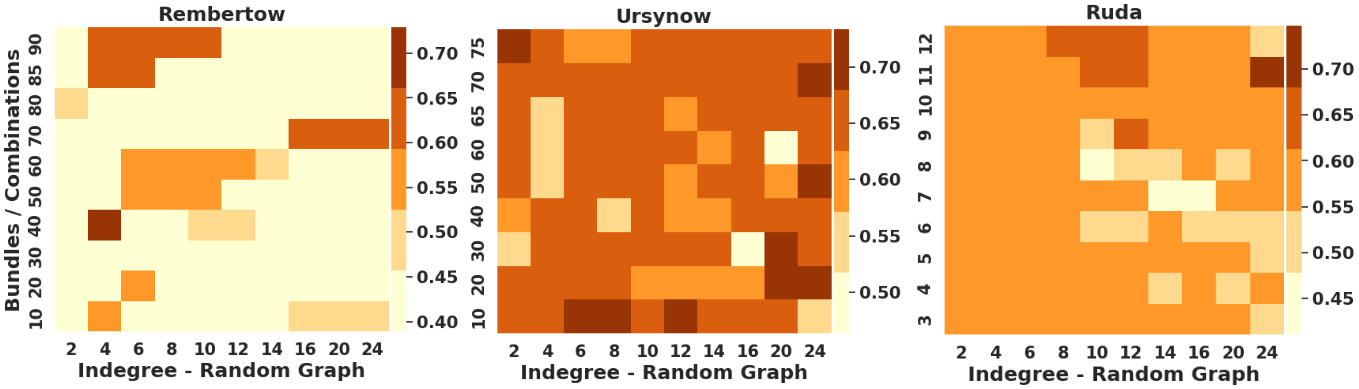}
    \caption{Popularity index (fitness of consensus) for different number of valid bundle combinations and in-degrees.}
    \label{fig:popul}
\end{figure}

Figure~\ref{fig:popul} shows the popularity index (fitness of the consensus) for different number of bundles and in-degrees.  In case of Ursynow, the consensus bundles have more projects and a higher percentage of popular projects (0.65 to 0.70), that also have medium costs, which results in higher overall budget utilization. The percentage of popular projects is low for Rembertow (0.40 to 0.45) and selected popular projects have considerably high costs too, as the overall budget utilization is also low (see Figure~\ref{fig:budget}).

{\em Comparison with other aggregation methods}: The overlap of projects between the consensus bundle (using the maximum number of bundles for action space) and the winners from the aggregation methods are calculated (see Table~\ref{tab:static4}). When a larger number of projects are listed, e.g. Rembertow, the overlap with consensus bundle is higher with equal shares (0.62) and Phragmen (0.61). Hence, these methods maximize fairness and representation and also produce more legitimate winners. For a lower number of projects, e.g. n Ruda, greedy has higher overlap (0.72) with the consensus bundle. The reward based iterative learning with communication can reach a consensus, which has a higher overlap with aggregation methods that promote fairness for a higher number of listed projects. 

\begin{table}[!htb]
\caption{
The overlap between the projects in the consensus bundle and three aggregation methods: greedy, equal shares and phragmen. The highest possible number of valid knapsack bundles are used. {\em G: Greedy, PG: Phragmen, ES$^*$: Equal Shares*, MARL-PB: Proposed approach}}
\centering
\begin{tabular}{llllllll}
    \hline
       Dataset & \multicolumn{4}{c}{Size of Consensus Bundle} & \multicolumn{3}{c}{MARL-PB Overlap}  \\ 
         \hline
        ~ & MARL-PB & ES & PG & G & ES & PG & G \\ \hline
        Rembertow & 8 & 9 & 9 & 6 & 0.62 & 0.61 & 0.49 \\ 
         Ursynow & 8 & 8 & 8 & 6 & 0.72 & 0.75 & 0.73 \\ 
        Ruda & 7 & 8 & 8 & 5 & 0.62 & 0.66 & 0.72 \\ \hline
\multicolumn{8}{l}{$*$ phragmen completion method used for the method of equal shares.}
      
         \\
    \end{tabular}
   \label{tab:static4}
\end{table}

\begin{figure}[!htb]
    \includegraphics[scale = 0.47]{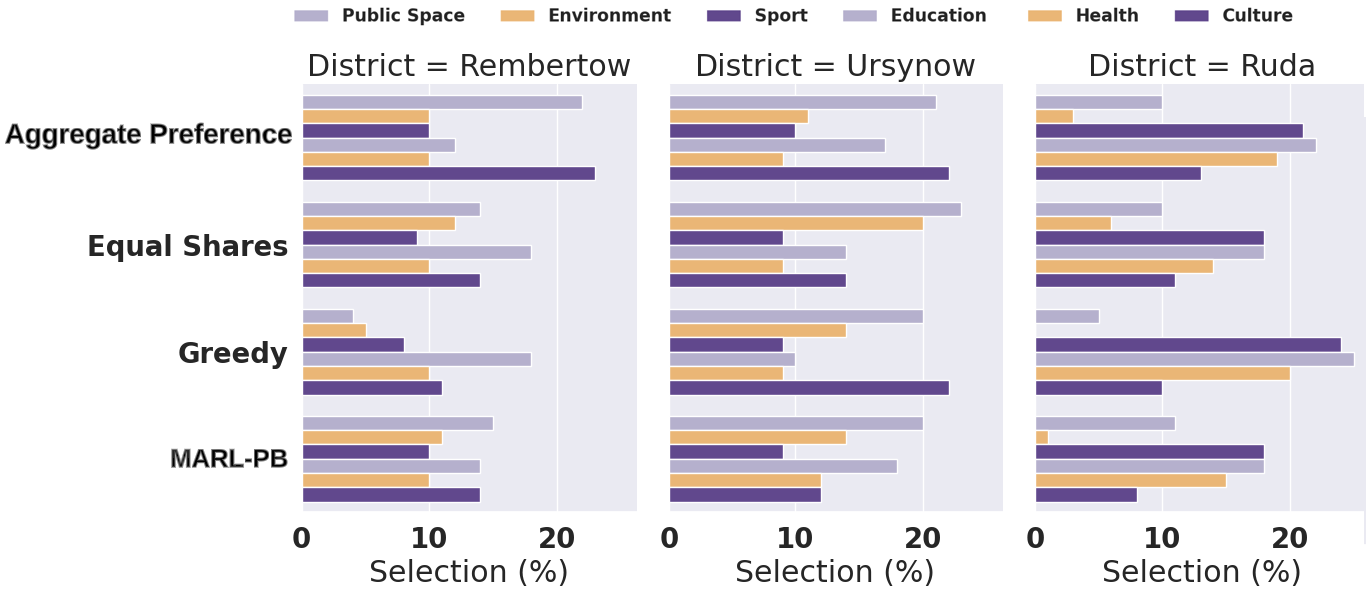}
    \caption{The attributes of the project selections with the different methods.}
    \label{fig:attr}
\end{figure}

{\em Analysis of the reward modelling}: The top-3 amenities associated with all projects (listed and selected) over all years in Ursynow are public space (22\%), education(17\%), environmental protection (12\%), impact on children (22\%) adults (21.1\%) and seniors (19\%). Similarly, for Rembertow and Ruda the most popular ones are public space (24\%) and education (22.7 \%). These project attributes affect a large proportion of the population. Figure~\ref{fig:attr} shows the amenities selected via MARL-PB and the aggregation methods, as well as how they compare with the original aggregate preferences based on the past data.  The projects selected using equal shares and MARL-PB correspond to similar public amenities (e.g. for Rembertow, projects related to public space, education and culture are selected in higher proportion). This also signifies that consensus projects in MARL-PB prioritize fairness and better representation. The public amenities selected in the greedy method do not correlate with the ones selected in the consensus for Rembertow and Ruda.  It can be observed for Ursynow that the selected projects for any aggregation method and with consensus mostly conform. Collective preferences for these projects remain stable over time in this region. 

\subsection{RQ2: Level of Flexibility to Compromise and Reach Consensus}

\begin{figure}[!htb]
    \centering
    \includegraphics[scale = 0.42]{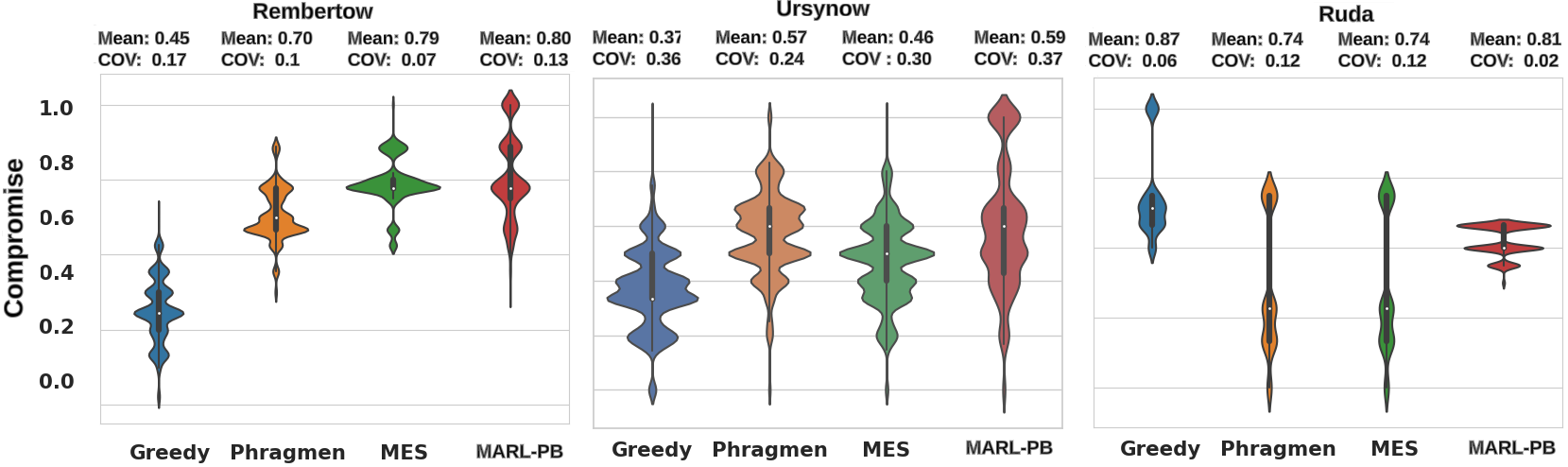}
    \caption{Mean voters' compromise cost of MARL-PB and the different voting aggregation methods. The coefficient of variation (COV) measures unfairness, which is how compromises spread within the voters' population.}
    \label{fig:sub123}
\end{figure}

Figure~\ref{fig:sub123} compares the the mean voters' compromise of MARL-PB with the one of different aggregation methods. The mean cost of compromise among the three districts for greedy is 0.56, while for MARL-PB is 0.68, which is close to the one of equal shares and Phragmen with 0.66 and 0.62 respectively. These results show the following: Consensus requires compromise that is not observed in the standard greedy voting aggregation method, however, this compromise is attainable and comparable to the one observed with consensus-oriented voting aggregation methods. 

The mean unfairness of MARL-PB is 0.17, while greedy is 0.19. The equal shares and Phragmen have an unfairness of 0.16 and 0.15 respectively. The case of Ruda does not align with the results of the other districts and this is likely an artifact of the lower number of projects.

\subsection{RQ3: Efficiency and Robustness}

\begin{figure}[!htb]
    \centering
    \includegraphics[scale = 0.74]{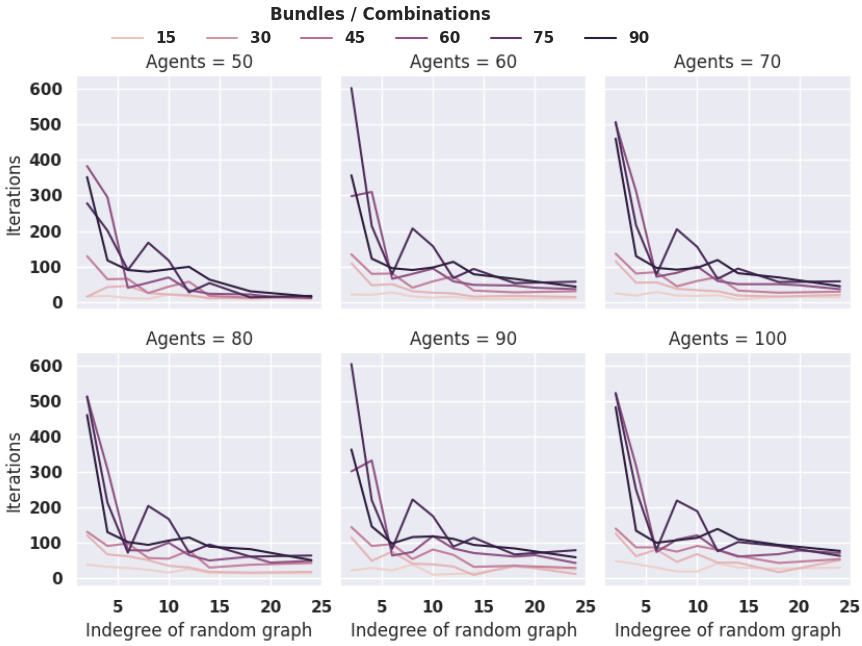}
 \caption{Convergence time increases as the number of agents increases. Significance values (p) from t-test (iterations, agents): Rembertow: p = 0.04, Ursynow: p = 0.04, Ruda: p = 0.04. The values here are averaged over the three districts.}
    \label{fig:agent}
\end{figure}

{\em Convergence}:  An increase in the number of agents increase the converge time for any combination of in-degree or action (bundle) space (see Figure~\ref{fig:agent}). The iterations increase by 5\% on average for an increase of 50 voters. Although this increase is only based on the data from the three districts, the system demonstrates to be scalable by converging within finite time as the number of voters increases.

\begin{figure}[!htb]
    \centering
    \includegraphics[scale = 0.41]{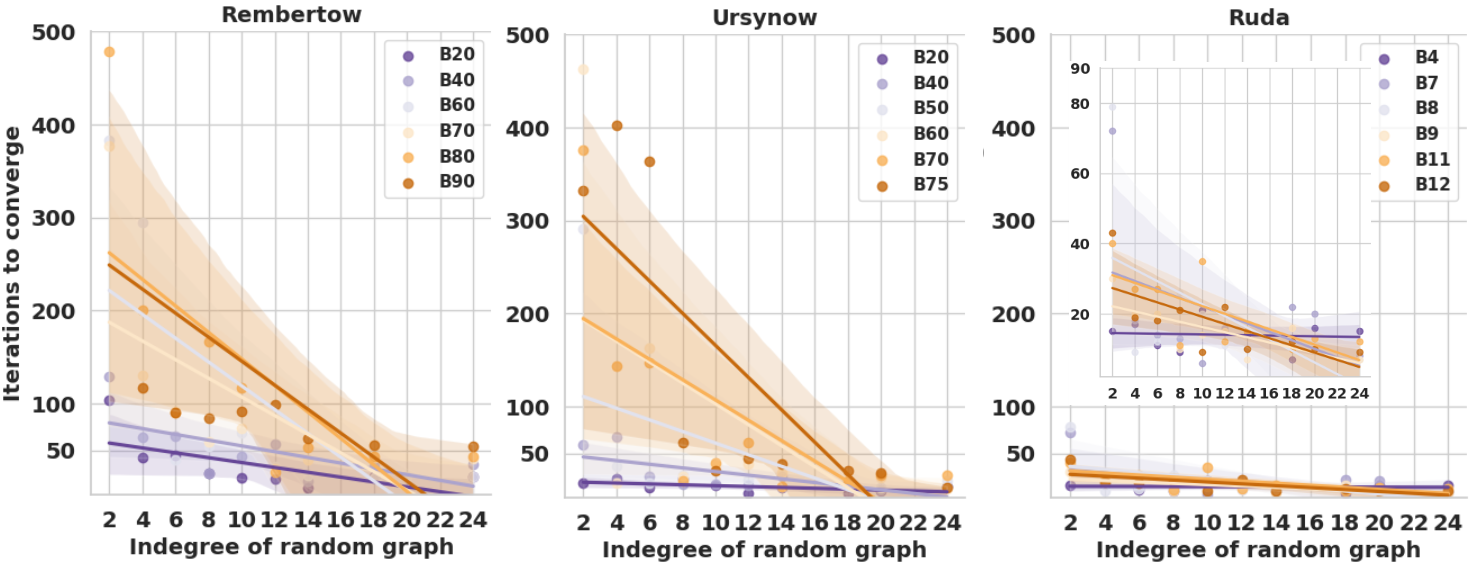}
    \caption{Convergence time decreases for smaller action (bundle) spaces and higher in-degrees: In-degree vs. iterations (significance values (p) from t-test): Rembertow: p = 0.001, Ursynow: p = 0.002, Ruda: p = 0.03). Bundle size vs. iterations (significance values (p) from t-test): Rembertow: p = 0.01, Ursynow: p = 0.001, Ruda: p = 0.04.}
    \label{fig:iter}
\end{figure}

Figure~\ref{fig:iter} shows for each district the convergence time as function of in-degree and bundles size. Smaller bundles with higher in-degrees improve the speed. 

\begin{figure}[!htb]
    \centering
    \includegraphics[scale = 0.43]{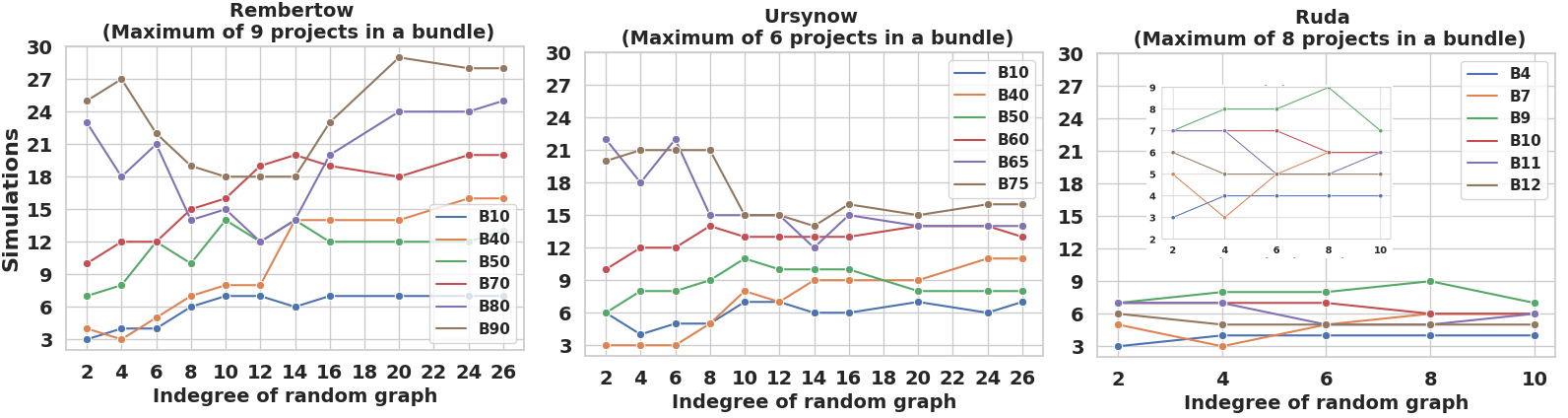}
    \caption{Number of repetitions (simulations) required to reach the same consensus for a certain set of parameters. In-degree vs. simulations (significance values (p) from t-test): Rembertow: p = 0.03, Ursynow: p = 0.03, Ruda: p = 0.04. Action (bundle) space vs. simulations (significance values (p) from t-test): Rembertow: p = 0.01, Ursynow: p = 0.02, Ruda: p = 0.04. The  number of projects in a bundle does not have significant impact on the simulations (Rembertow: p = 0.09, Ursynow: p = 0.06, Ruda: p = 0.13).}
    \label{fig:rob}
\end{figure}

{\em Robustness}: The influence of the randomness in the dynamic communication network on the stability of convergence is assessed by repeating the learning process multiple times, with different size of bundles. Figure~\ref{fig:rob} shows the required number of repetitions for a stable convergence speed. More repetitions are required for lower in-degrees due to limited  information exchange for deliberation. A higher action space (number of bundles) results in a larger number of alternatives to explore and thus stability requires a higher number of repetitions. 

\section{Conclusion and Future Work}\label{sec:Conclusion}

This paper concludes that a consensus-based participatory budgeting process, with three use cases introduced, is feasible via a novel multi-agent reinforcement learning approach. The consensus process actually models a more systematic, large-scale and automated deliberation process, which has so far remained decoupled from the collective choice of voting. The experimental evaluation with real-world data confirms that that the studied consensus is reachable, efficient and robust. The results also demonstrate that the consensus in MARL-PB requires compromises from voters, which are though comparable to the ones of existing voting aggregation methods that promote fairness and inclusion. 

This is a key result with impact and significant implications: voters may not need in the future to rely anymore on a top-down arbitrary selection of the aggregation method. Instead, communities will be empowered to institutionalize and directly apply independently their own consensus-based decision-making processes. Moreover, city authorities may use the proposed method to filter out projects during the project ideation phase, which usually relies on subjective criteria with risks on legitimacy. 

As part of future work, the agent communication may expand to different dynamic topologies that represent more closely social networks and proximity. The expansion of the multi-agent reinforcement learning approach with other preferential elicitation methods~\cite{hausladen2023legitimacy}, beyond approval voting, is expected to further strengthen the accuracy and legitimacy of consensus-based participatory budgeting. A more advanced design of the rewards scheme will further expand the applicability of this ambitious approach. 

\section*{Acknowledgements}\label{sec:Acknowledgment}

This work is supported by a UKRI Future Leaders Fellowship (MR\-/W009560\-/1): `\emph{Digitally Assisted Collective Governance of Smart City Commons--ARTIO}',  and the SNF NRP77 `Digital Transformation' project ``Digital Democracy: Innovations in Decision-making Processes", \#407740\_187249.

\bibliographystyle{splncs04}
\bibliography{ref}

\begin{thebibliography}{10}
\providecommand{\url}[1]{\texttt{#1}}
\providecommand{\urlprefix}{URL }
\providecommand{\doi}[1]{https://doi.org/#1}

\bibitem{airiau2017learning}
Airiau, S., Grandi, U., Perotto, F.S.: Learning agents for iterative voting.
  In: Algorithmic Decision Theory: 5th International Conference, ADT 2017,
  Luxembourg, Luxembourg, October 25--27, 2017, Proceedings 5. pp. 139--152.
  Springer (2017)

\bibitem{aziz2021participatory}
Aziz, H., Shah, N.: Participatory budgeting: Models and approaches. Pathways
  Between Social Science and Computational Social Science: Theories, Methods,
  and Interpretations pp. 215--236 (2021)

\bibitem{badanidiyuru2018bandits}
Badanidiyuru, A., Kleinberg, R., Slivkins, A.: Bandits with knapsacks. Journal
  of the ACM (JACM)  \textbf{65}(3),  1--55 (2018)

\bibitem{faliszewski2023participatory}
Faliszewski, P., Flis, J., Peters, D., Pierczy{\'n}ski, G., Skowron, P.,
  Stolicki, D., Szufa, S., Talmon, N.: Participatory budgeting: Data, tools,
  and analysis. arXiv preprint arXiv:2305.11035  (2023)

\bibitem{fletcher2018comparing}
Fletcher, S., Islam, M.Z., et~al.: Comparing sets of patterns with the jaccard
  index. Australasian Journal of Information Systems  \textbf{22} (2018)

\bibitem{ganuza2012deliberative}
Ganuza, E., Franc{\'e}s, F.: The deliberative turn in participation: the
  problem of inclusion and deliberative opportunities in participatory
  budgeting. European Political Science Review  \textbf{4}(2),  283--302 (2012)

\bibitem{hausladen2023legitimacy}
Hausladen, C.I., H{\"a}nggli, R., Helbing, D., Kunz, R., Wang, J., Pournaras,
  E.: On the legitimacy of voting methods. Available at SSRN 4372245  (2023)

\bibitem{jelasity2007gossip}
Jelasity, M., Voulgaris, S., Guerraoui, R., Kermarrec, A., Steen, M.:
  Gossip-based peer sampling. ACM Transactions on Computer Systems (TOCS)
  \textbf{25}(3) (2007)

\bibitem{kilgour2010approval}
Kilgour, D.M.: Approval balloting for multi-winner elections. In: Handbook on
  approval voting, pp. 105--124. Springer (2010)

\bibitem{Gilla2023}
Leibiker, G., Talmon, N.: {A Recommendation System for Participatory
  Budgeting}. In: International Conference on Autonomous Agents and Multiagent
  Systems (AAMAS). Workshop on Optimization and Learning in Multiagent Systems
  (2023)

\bibitem{liekah2019multiagent}
Liekah, L.: Multiagent reinforcement learning for iterative voting. Ph.D.
  thesis, Master’s thesis, Toulouse Capitole University (2019)

\bibitem{macy2002learning}
Macy, M.W., Flache, A.: Learning dynamics in social dilemmas. Proceedings of
  the National Academy of Sciences  \textbf{99}(suppl\_3),  7229--7236 (2002)

\bibitem{miller2019modes}
Miller, S.A., Hildreth, R., Stewart, L.M.: The modes of participation: A
  revised frame for identifying and analyzing participatory budgeting
  practices. Administration \& Society  \textbf{51}(8),  1254--1281 (2019)

\bibitem{slivkins2019introduction}
Slivkins, A., et~al.: Introduction to multi-armed bandits. Foundations and
  Trends in Machine Learning  \textbf{12}(1-2),  1--286 (2019)

\bibitem{stolicki2020pabulib}
Stolicki, D., Szufa, S., Talmon, N.: Pabulib: A participatory budgeting
  library. arXiv preprint arXiv:2012.06539  (2020)

\bibitem{wellings2023improving}
Wellings, T.S., Majumdar, S., Haenggli~Fricker, R., Pournaras, E.: Improving
  city life via legitimate and participatory policy-making: A data-driven
  approach in switzerland. In: Annual International Conference on Digital
  Government Research. pp. 23--35 (2023)

\end{thebibliography}

\end{document}